\documentclass[10pt,aps,prb,twocolumn,groupedaddress,superscriptaddress,
showpacs]{revtex4}

\usepackage{graphicx}
\usepackage{bm}
\usepackage{amsmath}
\usepackage{dcolumn}
\usepackage[usenames]{color}
\usepackage[normalem]{ulem}

\renewcommand{\figurename}{FIG.}
\makeatletter\renewcommand{\fnum@figure}[1]{\figurename~\thefigure.}\makeatother

\newcolumntype{.}{D{.}{.}{3}}  
\newcommand{\CMO}{CuMnO$_2$}
\newcommand{\CMOn}{Cu$_{1.04}$Mn$_{0.96}$O$_2$}

\newcommand{\mb}{\ensuremath{\mu_{\text{B}}}}

\begin{document}

\title{Orbital structure and magnetic ordering in stoichiometric and doped crednerite CuMnO$_2$}

\author{A. V. Ushakov}
\affiliation{Institute of metal physics, Russian academy of science, 
S. Kovalevskaya str. 18, 620041 Ekaterinburg, Russia}

\author{S. V. Streltsov}
\affiliation{Institute of metal physics, Russian academy of science, 
S. Kovalevskaya str. 18, 620041 Ekaterinburg, Russia}
\affiliation{Ural Federal University, Mira St. 19, 620002 Ekaterinburg, Russia}

\author{D. I. Khomskii}
\affiliation{II. Physikalisches Institut, Universit\"at zu K\"oln,
Z\"ulpicherstra{\ss}e 77, D-50937 K\"oln, Germany}

\begin{abstract}

The exchange interactions and magnetic structure in layered system \CMO ~(mineral crednerite) and in  
nonstoichiometric system \CMOn,
with triangular layers distorted due to orbital ordering of the Mn$^{3+}$ ions, 
are studied by {\it ab-initio} band-structure calculations, which were performed within the GGA+U 
approximation. The exchange interaction parameters for the Heisenberg model within the Mn-planes and between 
the Mn-planes are estimated.
We explain the observed in-plane magnetic structure by the dominant mechanism of the direct $d-d$ 
exchange between neighboring Mn ions. The superexchange via O ions, with 90$^{\circ}$ Mn-O-Mn bonds, 
plays less important role for the in-plane exchange. The interlayer coupling is largely 
dominated by one exchange path between the half-filled $3z^2-r^2$ orbitals of Mn$^{3+}$. The change of interlayer coupling from antiferromagnetic in pure \CMO~to ferromagnetic in doped material is also explained by our calculations.
\end{abstract}


\maketitle

\section{\label{sec:introd}Introduction}

Magnetic systems with geometric frustrations attract nowadays considerable attention. Due to competition of different exchange paths often rather complicated magnetic structures are realized, which are very sensitive to external influences and can be changed e.g. by small modifications of composition.

There exist many magnetic materials containing the triangular 
layers as a main building block.~\cite{Collins-97} Among them there are well-known systems like NaCoO$_2$,~\cite{Akimoto-03} 
delafossites $A$MeO$_2$ ($A$ = Cu, Ag, Pd, ...; Me -- transition metals), such as multiferroic CuFeO$_2$ or 
AgCrO$_2$,~\cite{Tokura-08,Kimura-10} compounds with unusual charge state like 
Ag$_2$NiO$_2$,~\cite{Johannes-07} ferroaxial multiferroic FeRb(MoO$_4$)$_2$~\cite{Kenzelmann-07} and even 
some organic systems with spin-liquid ground states.~\cite{Kanoda-05,Kanoda1-05} A regular triangular lattice 
with equilateral triangles is frustrated, 
which can lead to complicated ground states, especially in the presence of strong magnetic anisotropy. 
Such frustrations could in principle be lifted due to lattice distortion, for example 
caused by orbital ordering~\cite{Mostovoy-02,Mostovoy-03}(which, on the other hand, can itself lead to frustration in some, even not geometrically frustrated lattices~\cite{Khomskii-82,Mostovoy-03}). However in certain cases frustrations can remain even after such orbital ordering.

\begin{figure}[b!]
\begin{center}
\includegraphics[width=0.9\columnwidth]{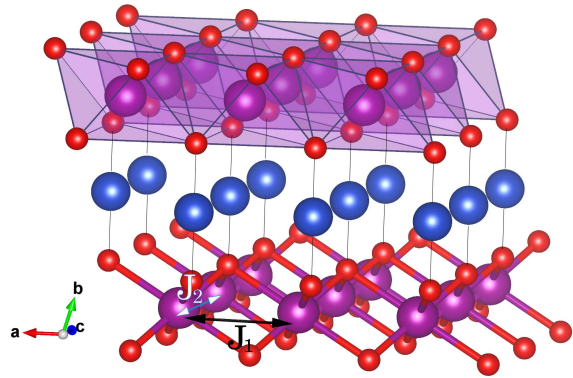}
\end{center}
\caption{\label{cryst.str} (Color online) The crystal structure of CuMnO$_2$ in the 
low temperature phase. Mn ions are shown as violet, Cu as blue, and O as 
red balls. Two different intralayer exchange constants via the long ($J_1$) and 
short ($J_2$) Mn-Mn bonds are shown. The MnO$_6$ layers alternating in $c$
direction are equivalent, but presented in slightly different ways to show
common edges of the MnO$_6$ octahedra (upper layer) and the Mn-O bonds
(lower layer). The image was generated using VESTA software.~\cite{Momma-11}
}
\end{figure}

\begin{figure}[tbp!]
\begin{center}
\includegraphics[width=0.9\columnwidth]{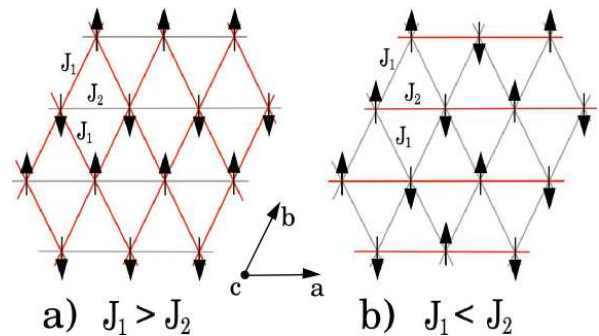}
\end{center}
\caption{\label{lattice} (Color online) Possible magnetic structures for the distorted triangular lattice. a) Exchange parameters $J_1$ on the long bonds are larger than that on short bonds $J_2$; b) Exchange parameters on long bonds smaller that on short ones. This structure corresponds to the observed in CuMnO$_2$ experimentally.~\cite{Damay-09} For both cases the strongest exchange paths are shown in red.}
\end{figure}

Apparently an example of this type is delafossite~\CMO~(mineral crednerite). The crystal
structure of CuMnO$_2$ is shown in Fig.~\ref{cryst.str}. Two features, disclosed by the experimental studies of stoichiometric \CMO~\cite{Damay-09, Martin-10} and similar system with the excess of Cu, \CMOn,~\cite{Chapon-11} are quite nontrivial and require explanation. First, the presence of the Jahn-Teller Mn$^{3+}$ ($t_{2g}^3e_g^1$) ions in this system leads to change of the crystal structure (with corresponding ferro--orbital ordering) from the usual for delafossites rhombohedral $R\overline{3}m$ to a monoclinic $C2/m$ structure already at room temperature. In this structure different directions in the triangular $ab$-plane become inequivalent: there exist for each Mn two short and four long Mn-Mn 
distances in the plane. This in principle could lift frustration, if the antiferromagnetic (AFM) 
exchange $J_1$ on the long Mn-Mn bonds would be stronger than the exchange $J_2$ on the short ones: 
then the topology of the plane would essentially become that of a square lattice Fig.~\ref{lattice}(a). 

However experimentally it turned out that this is not the case: apparently the 
exchange at short bonds is stronger, which, with the uniaxial magnetic anisotropy of 
Mn$^{3+}$ (spins are oriented predominantly along the long Mn-O bonds) still leaves 
the system frustrated [the stacking of the antiferromagnetic chains along the short 
Mn-Mn direction is frustrated for equivalent long bonds which couple such chains, 
see Fig.~\ref{lattice}(b)]. According to Ref.~\onlinecite{Damay-09} this degeneracy 
is lifted below magnetic transition: at $T<T_N=65$~K the structure changes from 
monoclinic to triclinic $C\overline{1}$ due to magnetostriction.

The first question is why indeed the exchange coupling along the short Mn-Mn bond is stronger. At first glance it seems very natural: exchange interaction is expected to be stronger for shorter bond. But, besides the direct Mn-Mn exchange due to the overlap of the $d$-orbitals of Mn,  usually also the superexchange via oxygen plays an important role, especially for the heavier $3d$ elements. And one could expect that for the ferro-orbital ordering like that found in CuMnO$_2$ this contribution could be stronger for longer Mn-(O)-Mn bonds (see in more details below). Why is this not the case, is a priori not clear.

Another surprising phenomenon was found in nonstoichiometric crednerite with small excess of copper, 
Cu$_{1.04}$Mn$_{0.96}$O$_2$, with part of the in-plane Mn substituted by Cu. It was found in 
Ref.~\onlinecite{Chapon-11} that, whereas the in-plane magnetic ordering (and magnetostrictive distortions) remain practically the same as for pure CuMnO$_2$, the interlayer exchange coupling changes to the opposite: if it was antiferromagnetic in CuMnO$_2$, it becomes ferromagnetic in Cu$_{1.04}$Mn$_{0.96}$O$_2$.~\cite{footnote} Interlayer coupling is not often treated seriously in the study of such layered frustrated systems; it is usually considered rather as a nuisance (although of course everyone understands that some such interlayer coupling is required to make a real 3$D$ long-range magnetic ordering). Here, however, we have to seriously address the question of such interlayer coupling and its dependence on the exact composition of the material.

To explain the observed features, one has to  know the values of exchange interaction between different 
Mn-Mn pairs, both in-plane and out-of-plane. For the lattice geometry at hand,  with its low symmetry and with 
many different exchange paths (exchange due to direct Mn-Mn overlap, different superexchange processes such 
as $t_{2g}$-$t_{2g}$, $t_{2g}$-$e_{g}$, $e_g$-$e_g$, see e.g. Ref.~\onlinecite{Streltsov-08}), especially in the presence of orbital ordering existing in CuMnO$_2$, it becomes a formidable task,  difficult to solve ``by hand''. Therefore we decided to address this problem by {\it ab initio} band-structure calculations, which automatically take into account all exchange mechanisms, different bond lengths and angles present in this system.
This allows us to compare relative stability of different types of magnetic ordering, for both stoichiometric and for doped
 CuMnO$_2$.
To avoid symmetry changing of the system we did not use the supercell calculations with some 
particular Mn ions substituted by Cu to model \CMOn, but modeled the doped situation by changing 
the electron concentration, which is equivalent to virtual crystal approximation (VCA).

The results of our calculations, first of all, confirm that the experimentally observed magnetic structure is the theoretical ground state and also confirm the results of the previous first principle calculations~\cite{Jia-12} for the stoichiometric case. 
The analysis shows, that the hopping between the $3z^2-r^2$ and $xy$ orbitals and 
the direct hopping between $xy$ orbitals  should give the main 
contribution to the $J_2$ exchange, and this AFM exchange turns out to 
be much stronger than $J_1$, leading to the magnetic structure observed experimentally. 
The interlayer ordering from the calculations also coincides with that observed experimentally. 
For the nonstoichiometric CuMnO$_2$ the in-plane ordering remains the same, whereas the 
interlayer ordering changes to the opposite, in agreement with the findings of Ref.~\onlinecite{Chapon-11}. 
We discuss the physical mechanism of this change.

\section{Calculation details}

The calculations were carried out by pseudopotential method in the PWscf code~\cite{PW-SCF} in the 
framework of the density functional theory (DFT). We used the Vanderbilt ultrasoft 
pseudopotentials for all ions. $3d$, $4s$, $4p$ for Cu and Mn, $2s$ and $2p$ for O were
considered as valence.
The plane-wave cutoff energy for these pseudopotentials was taken be 40 Ry.
The strong electronic correlations were included on Mn sites via the GGA+$U$ approximation~\cite{Anisimov-97}. 
Effective $U_{eff}=U-J_H$ for the Mn 3$d$ states was chosen to be 4.1 eV.~\cite{Wang2006}
We checked that main conclusions doesn't depend on the choice of $U_{eff}$ and also 
present the results for $U_{eff}=$3.6~eV.

The integration in the course of self-consistency iterations was performed
over a mesh of 256 $k$-points in a whole part of the Brillouin zone with switching off the symmetry of 
the space group to allow any possible orbital order. We used the Perdew-Burke-Ernzerhof (PBE) 
exchange-correlation functional.~\cite{Perdew-96}

In order to estimate the exchange interaction parameters within the Mn triangular layers and 
the interlayer exchange we calculated the total energies of different magnetic configurations 
and used the Heisenberg model in the form 
\begin{equation}
\label{heis}
H = \sum_{ij} J_{ij} S_iS_j,
\end{equation}
where summation runs twice over each pair. The in-plane exchange constants $J_1$ and $J_2$ 
(see Fig.~\ref{lattice}) were recalculated from the total energies
of three magnetic structures: (1) fully ferromagnetic,
(2) Mn ions along the short (long) Mn--Mn bond are ordered ferromagnetically 
(antiferromagnetically), (3) Mn ions along the short Mn--Mn bonds are 
antiferromagnetically coupled. The unit cell contains 4 f.u. In the LT phase the exchange 
parameters $J_1$ along two long Mn--Mn bonds were assumed to be the same,
since the difference of the bond lengths is quite small, 
$\sim 0.02 \AA$.  For the calculation of the exchange constants
along the $b-$axis the unit cell was doubled in this direction, the magnetic ordering 
in plane was chosen as in the experiment (ferromagnetic chains are going along the longest 
Mn--Mn bond).

The lattice parameters (unit cell; atomic coordinates) used in the calculations were taken from the 
experimental data of Ref.~\onlinecite{Damay-09,Martin-10,Chapon-11} for $300 K$ (HT phase) and 
for $10 K$ (LT phase).

\section{Results and discussion}
The first principles calculations of the stoichiometric CuMnO$_2$ were performed for the 
high-temperature (HT) and low-temperature (LT) phases. The magnetic moment on Mn$^{3+}$($d^4$) 
is close $3.7$~$\mb$ in both phases (3.68~$\mb$ in LT and 3.74~$\mb$ in HT phase). The 
occupancy of the $3d$ shell for the interlayer linearly-coordinated Cu$^{1+}$ was
found to be 9.73 in LT and 9.71 in HT phase. This is close to the $d^{10}$ configuration
expected from the simple ionic consideration. The Mn$^{3+}$ ion has $5.58$ (5.54) electrons 
in the LT (HT) phases. The deviation from the nominal filling $d^4$ is related
with the hybridization effects as discussed below.

According to the GGA+U calculations CuMnO$_2$ is a metal at the HT phase in the ferromagnetic configuration, while at lower temperatures in the experimental~\cite{Damay-09} AFM 
magnetic structure
the energy gap opens, $E_g = 0.2$ eV, and system becomes insulating. We found that such magnetic order, 
with the AFM stripes in the Mn-plane and the AFM arrangement along the crystallographic $b$-axis [Fig.~\ref{lattice}(b)]
is the ground state in the LT phase with the lowest total energy.

\begin{figure}[tbp!]
\begin{center}
\includegraphics[width=0.8\columnwidth,angle=270]{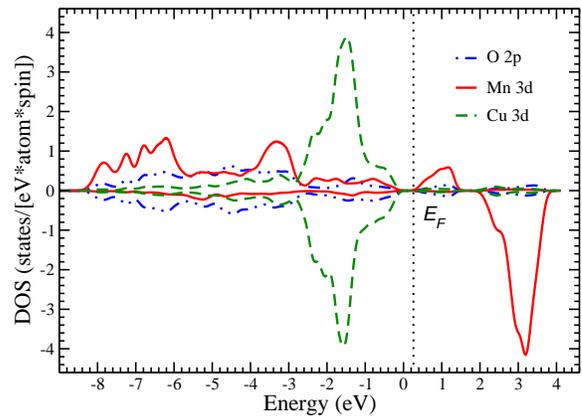}
\end{center}
\caption{\label{total-dos} (Color online) The partial densities of states (DOS) obtained in the GGA+$U$ calculation for 
stoichiometric CuMnO$_2$ at LT phase in the AFM configuration, shown in Fig.~\ref{lattice}(b). 
Positive (negative) values correspond to spin up (down). The Fermi energy is in zero.}
\end{figure}

The partial Cu $3d$, Mn $3d$ and O $2p$ densities of states for the LT phase are presented in Fig.~\ref{total-dos}. 
One can see, that the Mn $d$ states are lower and broader than the Cu $3d$ ones, which are 
closer to the Fermi level. A substantial hybridization between the Mn $3d$ and O $2p$ states 
is noticeable. Because of the hybridization the occupation of the $3d$ shell of Mn is larger than the 
nominal, but the spin moment is consistent with the $d^4$ configuration ($3.7 \mu_B$).

The structural distortions caused by the Jahn-Teller character of the Mn$^{3+}$
and orbital ordering results in the structure with two types of inequivalent Mn-Mn pairs 
in the MnO$_2$ plane: those along the $a$-direction with the short Mn-Mn distance of $2.88$~\AA,
and the long bonds in two other directions, with the Mn-Mn distance of $\sim 3.14$~\AA (they become 
inequivalent in the magnetically-ordered phase), see Fig.~\ref{lattice}. In effect one may expect that the exchange
coupling in the $ac$ plane can be characterized by two exchange constants: $J_2$ (short Mn-Mn bonds) 
and $J_1$ (long Mn-Mn bonds). The direct calculation in the GGA+U approximation shows that 
$J_1 =-1.5 K$ (FM) and $J_2 = 16.5 K$ (AFM).

\begin{figure}[tbp!]
\begin{center}
\includegraphics[width=0.9\columnwidth]{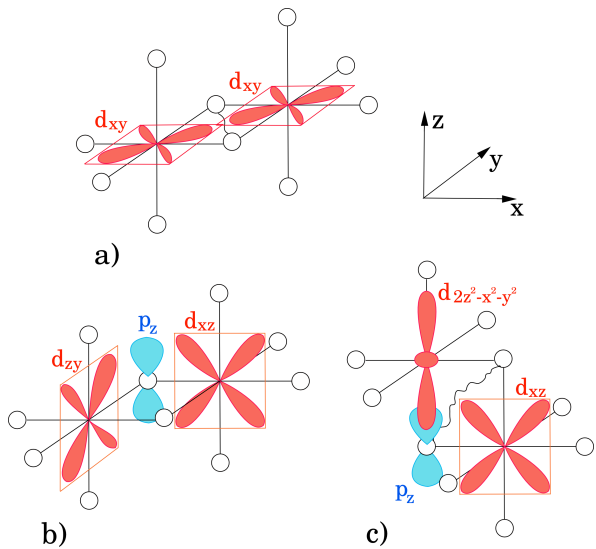}
\end{center}
\caption{\label{pathes} (Color online) Different processes for the exchange coupling of the neighboring 
Mn$^{3+}$ ions in the basal plane for the MnO$_6$ octahedra with common edge. (a) Direct overlap of the $t_{2g}$ orbitals; 
(b) antiferromagnetic $t_{2g}$--$t_{2g}$ superexchange via oxygen;
(c) antiferromagnetic $t_{2g}$--$e_g$ superexchange via oxygen; the Jahn-Teller distortion makes the distance 
between these two Mn ions long. Common edge of the neighboring MnO$_6$ octahedra 
is marked by wavy line. Oxygen ions are shown by circles. }
\end{figure}

The MnO$_6$ octahedra in the basal triangular layer have common edge, so that the Mn-O-Mn angle is close to $90^{\circ}$. There exist in 
this geometry several contributions to the nearest neighbor magnetic coupling. First, there is a 
direct hopping between the Mn orbitals, especially $t_{2g}$ ones, with the lobes directed toward one another, 
see Fig.~\ref{pathes}(a). This contribution must be antiferromagnetic. The direct exchange
is known to be quite efficient for first few $3d$ transition metal ions, but it 
gradually decreases going from left to the right in the periodic table, and 
already for Cr and Mn the superexchange may become larger.~\cite{Streltsov-08}

Let us consider contributions to the superexchange via O ions, some of which are illustrated in 
Fig.~\ref{pathes}(b,c) (see more details e.g. in Ref.~\onlinecite{Streltsov-08}).
For a such Mn-O-Mn $90^{\circ}$ bond, according to Goodenough-Kanamori-Anderson rules (GKA),~\cite{Khomskii-01,Goodenough} 
the superexchange between the half-filled $t_{2g}$ orbitals via oxygens will be 
also antiferromagnetic $\sim \frac{t_{pd\pi}^4}{\Delta^2} (\frac{1}{U}+\frac{2}{2\Delta + U_{pp}})$, where $\Delta$ is the 
charge-transfer energy (energy of the excitation, in our case, 
Mn$^{3+}$($d^4$)O$^{2-}$(2$p^6$) $\rightarrow$ Mn$^{2+}$($d^5$) O$^-$(2$p^5$)), and $U_{pp}$ is the repulsion of oxygen $p$-electrons.

But potentially the most important contribution to the Mn-Mn exchange via oxygens could be 
the $t_{2g}$-$e_g$ exchange, which, for the hopping to the half-filled $e_g$ orbital, is antiferromagnetic, 
and could be quite strong, $\sim \frac{t_{pd\pi}^2t_{pd\sigma}^2}{\Delta^2}(\frac{1}{U} + \frac{2}{2\Delta + U_{pp}})$. 
In pair of the MnO$_6$ octahedra forming long Mn-Mn bond the single half-filled $e_g$ 
orbital ($2z^2-x^2-y^2$) will lie in the plane  defined by common edge and two Mn ions
and will take part in such an exchange process as shown in Fig.~\ref{pathes}(c). Thus this
contribution will be stronger for longer Mn-Mn bond. 
This is what we had in mind in the Sec.~\ref{sec:introd}, where mentioned
that superexchange may leads to the situation when $J_1$ would be stronger
than $J_2$. But this not the case and direct exchange obviously
dominates.  
 
It has to be mentioned again that actually, it is a priori not clear whether -- the 
direct $d-d$ exchange on a short Mn-Mn bond or the 
superexchange via oxygens on the long bonds would be stronger. In the second case we 
would have expected that the magnetic ordering would be simple two-sublattice 
antiferromagnetism in an effective square lattice formed by the long Mn-Mn bonds, 
Fig.~\ref{lattice}(a) (it can also be viewed as the stripe ordering in the original triangular lattice, but with stripes of parallel spins running along short bonds, not along long ones, as in Fig.~\ref{lattice}(b)). Note that such a situation is indeed realized on some triangular lattiсes,  e.g. in NaVO$_2$~\cite{Cava-08} (in this case the orbital 
ordering and the direct $d$--$d$ exchange relieve geometric frustration).

There exist also other contributions to the nearest neighbor exchange, e.g. those involving one 
occupied and one empty orbital; according to the GKA rules, these processes would give ferromagnetic contribution, but weaker by the factor of $\frac{J_{H}}{U}$ or $\frac{J_{H}}{2\Delta}$. Still, there are many such exchange channels, so that the total contribution of these processes can be significant. 

The results of our {\it ab initio} calculations give an answer to the question raised above; 
it turns out that the antiferromagnetic exchange $J_2$ on the short Mn--Mn bond is stronger 
than the superexchange contributions on the long bonds. Apparently the rather short Mn--Mn 
distance ($2.88$~\AA) on a short bond makes direct $t_{2g}$--$t_{2g}$ exchange dominant,
but the superexchange between $e_g$ and $t_{2g}$ orbitals via oxygen should
also be important.

\begin{figure}[tbp!]
\begin{center}
\includegraphics[width=0.9\columnwidth]{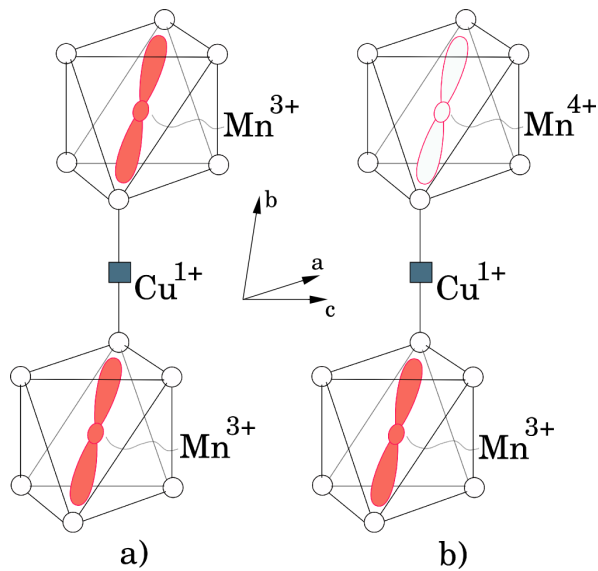}
\end{center}
\caption{\label{beetwen} (Color online) a) The strongest exchange path for interlayer
          antiferromagnetic coupling (involving the half-filled $e_g$-orbitals) in the stoichiometric \CMO. Oxygens are shown by circles, Cu$^{1+}$($d^{10}$)
         ion is shown by square.
         b) The same exchange path, but for the upper Mn$^{4+}$ (for the nonstoichiometric Cu$_{1+x}$Mn$_{1-x}$O$_2$); the hopping would be to the empty $e_g$-orbital which is not shaded. According to the GKA rules this exchange would be ferromagnetic. }
\end{figure}

In order to estimate the interlayer exchange interaction parameters and also to 
simulate the doped system Cu$_{1+x}$Mn$_{1-x}$O$_2$ we used the supercell with $8$ 
inequivalent Mn ions. Two different magnetic configurations were calculated.
In both structures Mn ions constituting short in-plane Mn--Mn bonds were 
antiferromagnetically coupled, while the interlayer spin order was taken different, ferro- or antiferromagnetic.~\cite{footnote} The calculated exchange 
constant $J_{inter} = 0.8 K$ (AFM) reproduced the observed interlayer ordering for undoped 
CuMnO$_2$.~\cite{Martin-10,Jia-12} the main contribution to the interlayer exchange is 
given by the process illustrated in Fig.~\ref{beetwen}.

The strongest exchange path goes from the occupied (half-filled) $3z^2-r^2$ orbital of Mn ion in one layer, where the local $z$-axis is directed along the long Mn-O bond, via corresponding oxygen $2p$ orbitals 
and eventually diamagnetic Cu$^{1+}$ sitting in between layers, and then to the similar $3z^2-r^2$ orbital of one particular 
Mn in the next layer (see Fig.~\ref{beetwen}(a)). This exchange coupling is antiferromagnetic, which provides the antiferromagnetic 
coupling between layers observed for CuMnO$_2$.~\cite{Damay-09} Note that the Mn ions connected by this exchange path 
do not belong to one unit cell, i.e. one must be careful in comparing the obtained magnetic ordering with the experimental one (which is defined in Ref.~\onlinecite{Damay-09} in terms of relative orientation of crystallographically equivalent Mn ions in neighboring layers).

As it was mentioned above the same (as for pure CuMnO$_2$) supercell consisting of the 
8 Mn ions was used to simulate the magnetic properties of the nonstoichiometric Cu$_{1+x}$Mn$_{1-x}$O$_2$ 
with x=0.04. Since the exact positions of the  doped Cu$^{2+}$ in the Mn$^{3+}$-plane is unknown, the virtual 
crystal approximation was used: an extra $0.32$ holes were added in the calculations of the 
aforementioned superсell. This corresponds to the uniform distribution of this hole over a whole cell.

The occupations of the $3d$ shell of the Cu and Mn ions in doped system are slightly different from
the stoichiometric case, being 9.71 and 5.55 respectively.
The magnetic moment on Mn in triangular plane is $3.69$ \mb. 
The occupations for different $3d$ orbitals are the same as in pure CuMnO$_2$.
The interlayer exchange for $x=0.04$ was found to be $J_{inter} = -1.8 K$, ferromagnetic, instead of antiferromagnetic interlayer coupling of $+0.8 K$ for undoped 
compound. The intralayer exchanges are $J_1=-6.4$~K and $J_2=-14.6$~K. 
Thus our calculations fully 
reproduce experimentally observed changes of the sign for the 
interlayer exchange coupling.~\cite{Chapon-11} This can be explained in the following way: when we substitute some Mn ions by Cu$^{2+}$ (formally trivalent Cu$^{3+}$ is rather difficult to obtain, and it cannot be formed at these conditions, which is confirmed by our calculations), we induce two changes. One is that 
Cu$^{2+}$ itself is magnetic and has different orbital occupation, 
so that for some ions the same exchange path from the $3z^2-r^2$ orbital of the Mn in the lower plane would connect to 
Cu$^{2+}$ in the neighboring plane, for which the $3z^2-r^2$ orbital will be completely filled. The $d$-hole will 
be, as always, on $x^2-y^2$ orbital. Due to different orbital occupation 
this exchange would be ferromagnetic, in accordance with the GKA rules.

Another consequence of the substitution of Mn$^{3+}$ by Cu$^{2+}$ is that to guarantee electroneutrality 
one Mn ion per each Cu should become Mn$^{4+}$. These Mn$^{4+}$ ions, e.g. in the upper layer,  would couple 
to Mn$^{3+}$ in the lower layer also ferromagnetically, see Fig.~\ref{beetwen}(b). Besides, as just explained, 
each Cu$^{2+}$ in triangular layer, as well as each Mn$^{4+}$, would couple ferromagnetically to both the 
layer above and layer below. Thus effectively each extra Cu$^{2+}$ would make {\it four} interlayer bonds 
ferromagnetic instead of antiferromagnetic. Apparently all these factor combine to lead to the inversion of 
the interlayer magnetic ordering in nonstoichiometric crednerite with the excess of Cu. And the fact that already 
very small amount of excess copper, only $4\%$ in Cu$_{1.04}$Mn$_{0.96}$O$_2$ studied in Ref.~\onlinecite{Chapon-11}, 
is sufficient to lead to this inversion of interlayer ordering, is probably connected with the factors discussed 
above: that effectively every extra Cu changes to the opposite the exchange of four  interlayer bonds, so that 
the effective doping is not $4\%$, but is rather $\sim 16\%$ (i.e. $16\%$ of strongest interlayer bonds change sign).

There may be also other factors contributing to the same effect. Notably, it is known, e.g. on the example of 
CMR manganites, that the substitution of Mn by other ions with different valence can modify the in-plane 
magnetic ordering in a rather large region around the dopant.~\cite{Raveau-01} It is not excluded that 
also here the in-plane ordering could be modified close to Cu$^{2+}$ and especially to Mn$^{4+}$ which should 
be created simultaneously. The interaction of this distorted region in a given plane with the next plane 
with its, say, original ordering could also have opposite sign from the interaction of two ``virgin'' planes. 
The eventual presence of magnetic distortions due to doping in CuMnO$_2$ should be checked by special 
experiments (the simplest indication of that would be certain broadening of magnetic reflexes in nonstoichiometric crednerite as compared to those in pure CuMnO$_2$ or ESR measurements).

It's worthwhile mentioning that the results obtained in the present investigation
does not strongly depend on the value on site Hubbard repulsion $U_{eff}$.
In order to check $U$ dependence we performed additional calculations
with smaller $U_{eff}=3.6$~eV. The decrease of the $U_{eff}$ leads
to increase of the AFM contributions to the intralayer exchange
coupling, which are known to be $\sim t^2/U$. As a result
both in-plane exchanges become more AFM: $J_1$ equals 21.0~K and 17.2~K, 
while $J_2$ is $-0.2$~K and $-5.1$~K for pure CuMnO$_2$ and nonstoichiometric 
case respectively. For $U_{eff}=3.6$~eV the interlayer exchange coupling changes its
sign going from CuMnO$_2$ ($J_{inter}=0.6$~K) to Cu$_{1.04}$Mn$_{0.96}$O$_2$
($J_{inter}=-1.1$~K) as for $U_{eff}$=4.1~eV.

The results obtained in the present paper agree with the conclusions of Ref.~\onlinecite{Terada2013}, 
where it was shown that the magnetic properties of CuMnO$_2$ strongly depends of what kind of dopant 
is used: magnetic Cu$^{2+}$ with not completely filled $3d$ shell or non-magnetic Ga$^{3+}$. In the 
first case the interlayer coupling changes to the opposite, while in the second in remains the same 
(antiferromagnetic). This demonstrates that this change is not due to a modification of the crystal
structure, but is connected with the substitution of Mn$^{3+}$ by magnetic Cu$^{2+}$,  with the 
generation of another  magnetic ion Mn$^{4+}$, as explained above.

\section{Conclusion}
In conclusion, on the basis of {\it ab initio} band structure calculations we obtained the physical picture,
which explains experimentally observed stripy antiferromagnetic order in stoichiometric crednerite CuMnO$_2$, 
as well as in the system with the excess of Cu, Cu$_{1+x}$Mn$_{1-x}$O$_2$. Ferro-orbital ordering present 
in this system plays very important role in determining the exchange constants and finally the magnetic structure. 
We argue that the in-plane magnetic ordering is mainly provided by the direct exchange interaction between the 
$t_{2g}$, while superexchange between $e_g$ and $t_{2g}$ orbitals on different sites is 
expected to be smaller. 
The interlayer exchange is mainly given by the exchange path involving the half-filled $e_g$  orbitals of
Mn$^{3+}$. Substitution of a part of Mn$^{3+}$ by Cu$^{2+}$, with corresponding creation of compensating 
Mn$^{4+}$ ions, leads to the inversion of the interlayer coupling already for rather small doping, which explains the puzzling experimental 
observation of Poienar et al. in Ref.~\onlinecite{Chapon-11} 

The obtained results once again  demonstrate the importance of orbital ordering for magnetic structures, this time in a frustrated system. It also shows that such systems are very sensitive to even small variations of conditions, e.g. small doping, and their properties can be effectively modified even by small perturbations. 

\section{Acknowledgments}
The authors are very grateful to Christine Martin for attracting our attention to this problem 
and for numerous discussions. This work was supported by the German projects FOR $1346$, by the University 
of Cologne via German Excellence Initiative and by the European project SOPRANO, Samsung by GRO program, 
and by the Russian Foundation
for Basic Research via RFFI-$13$-$02$-$00374$ of the Ministry of education and science of Russia 
(grant MK-$3443.2013.2$) .

\bibliography{./CuMnO2}

\begin{thebibliography}{27}
\bibitem[{foo()}]{footnote}
\bibinfo{note}{These notions require some explanation, they are not
  self-evident for triangular lattices with antiferromagnetic ordering in the
  layer. By antiferro- or ferro interlayer ordering one has in mind orientation
  of Mn ions in a corresponding unit cell, lying in different layers and
  connected by the lattice vector c = ($001$).}
\end{thebibliography}

\end{document}